\documentclass{ws-spin}
\usepackage{multicol}
\usepackage{xcolor}
\usepackage{cite}

\begin{document}

\markboth{M.~B.~Jungfleisch et al.}{New pathways towards efficient metallic spin Hall spintronics}

\title{New pathways towards efficient metallic spin Hall spintronics}

\author{Matthias Benjamin Jungfleisch}

\address{Materials Science Division, Argonne National Laboratory,\\
Argonne, Illinois 60439, USA\\
\email{jungfleisch@anl.gov}
}

\author{Wei Zhang}
\address{Materials Science Division, Argonne National Laboratory,\\
Argonne, Illinois 60439, USA\\
\email{zwei@anl.gov}
}

\author{Wanjun Jiang}
\address{Materials Science Division, Argonne National Laboratory,\\
Argonne, Illinois 60439, USA\\
\email{jiangw@anl.gov}
}

\author{Axel Hoffmann}
\address{Materials Science Division, Argonne National Laboratory,\\
Argonne, Illinois 60439, USA\\
\email{hoffmann@anl.gov}
}
\maketitle

\begin{history}
\end{history}

\begin{abstract}
Spin Hall effects interconvert spin- and charge currents 
due to spin-orbit interaction, which enables convenient electrical generation and detection of diffusive spin currents and even collective spin excitations in magnetic solids. Here, we review recent experimental efforts exploring efficient spin Hall detector materials as well as new approaches to drive collective magnetization dynamics and to manipulate spin textures by spin Hall effects. These studies are also expected to impact practical spintronics applications beyond their significance in fundamental research.

\end{abstract}


\begin{multicols}{2}

\section{Introduction}
\label{intro}
Spintronics relies not only on the charge degree of freedom, but also utilizes the spin degree of freedom and enables the manipulation of material properties and control devices. Towards this end, creation, detection and manipulation of spin currents are at the heart of spintronics. Historically, the discovery of giant magnetoresistance set a milestone in spintronics research. Soon after its discovery, the field contributed significantly to the transformation of modern information technologies, in particular with the development of high density magnetic recording and new concepts for non-volatile solid state memory devices \protect\cite{Hoffmann_PRA_2015}. Although, spin Hall effects (SHE) were proposed by D'yakonov and Perel' over four decades ago, substantial experimental efforts only started recently\protect\cite{Hoffmann_PRA_2015,Hoffmann_IEEE_2013,Saitoh_APL_2006,Kajiwara_Nat_2010,Mosendz_PRB_2010,Wang_PRL_2014,Zhang_PRL_2014,Jungfleisch_PRB_2015,Hahn_PRB_2013,Kelly_APL_2013,Wang_PRL_2014_anti,Du_JAP_2015,Sandweg_PRL_2011,Chumak_APL_2012,Jungfleisch_APL_2013,Jungfleisch_APL_2011,Zhang_PRB_2015,Weiler_PRL_2013,Weiler_PRL_2014} and within a short period of time, spin Hall effects have evolved from an academic curiosity to real applications (e.g., spin current detection, excitation of spin dynamics and magnetization switching).

This article will review recent experimental research efforts towards new, metallic spin-Hall phenomena and spintronics applications. Section~\ref{intro} will introduce basic theoretical concepts related to spin Hall effect physics including spin pumping and spin Seebeck effect. Following this, in Section~\ref{materials}, we will discuss novel spin Hall detector materials in light of high efficiencies of spin-to-charge conversion. Section~\ref{collective} will present new approaches to drive collective spin dynamics by means of SHEs in heterostructures consisting of a normal metal in contact with a  magnetic material. Lastly, in Section~\ref{textures}, we will focus on the manipulation of topologically stable spin textures by SHEs. 

\subsection{Spin Hall effects}
In 1879, Edwin E. Hall discovered the effect that was named after him \cite{Hall}. The Hall effect describes the generation of a transverse voltage in a conductor in the presence of a perpendicular magnetic field due to the Lorentz force on the electric charge current. In ferromagnetic metals, however, the voltage is not directly proportional to the magnetic filed. It also depends on the magnetization and this effect is known as the anomalous Hall effect \cite{Anomalous_Hall,Jiang_PRB_2010}. Depending on their spin orientation, electrons flowing in a ferromagnetic conductor will acquire different 
transverse velocities and since the electric charge current in a ferromagnet is polarized, the spin-dependent velocity leads to a transverse voltage.
The SHE is closely related to the anomalous Hall effect since the development of the spin-dependent transverse velocities is also present in non-magnetic materials. If we consider an electric charge current flowing through a non-ferromagnetic material, such as a paramagnetic metal or a doped semiconductor, electrons will experience an asymmetric scattering depending on the orientation of their spin: Electrons with spin-up will be scattered preferably in one direction perpendicular to the flow of the electric charge current and electrons with spin-down in the opposite direction \cite{Hoffmann_IEEE_2013,Hirsch_PRL_1999}. This spin current evolving transversally to the charge current with a spin orientation perpendicular to the plane, spanned by the two currents, is called SHE \cite{Hoffmann_IEEE_2013,Hirsch_PRL_1999,Dyakonov_JETP_1971,Dyakonov_PLA_1971,Zhang_PRL_2000}. The generation of charge current due to a gradient in the spin accumulation is called inverse spin Hall effect (ISHE)\cite{Saitoh_APL_2006}. The conversion from a spin- into charge current is described by:
\begin{equation}
\vec{J}_\mathrm{C}\propto\theta_\mathrm{SH}\vec{J}_\mathrm{S}\times\vec{\sigma}.
\label{eq2}
\end{equation}
Here, $\vec{J}_\mathrm{C}$ is the charge-current density, $\theta_\mathrm{SH}$ is the spin Hall angle, 
$\vec{J}_\mathrm{S}$ is the spin-current density and $\vec{\sigma}$ is the spin-polarization vector. The spin Hall angle is a material-specific parameter and the absolute value and the polarity of $\theta_\mathrm{SH}$ depends on the band structure of the spin Hall detector material \cite{Guo_PRL_2008}.

There are mainly three possible contributions to anomalous and SHEs \cite{Hoffmann_IEEE_2013}: (1) Skew scattering describes the asymmetric scattering of spin-up and spin-down electrons on a spherical potential due to the spin-orbit interaction which leads to different final momenta. (2) Side jump is caused by a spin-dependent difference in acceleration and deceleration during scattering. This leads to a finite spin-dependent displacement of the unpolarized charge current, and (3) intrinsic effects, which are given by the electronic band structure and due to a combination of spin-orbit coupling and virtual interband transitions that result in spin-dependent transverse velocities. 

As shown in Fig.~\ref{fig1} the spin Hall angle can either be positive ore negative. Theoretical models show that the intrinsic spin Hall conductivity depends on the spin-orbit polarization at the Fermi level, which is positive in metals with more than half filling and negative for less than half filling \cite{Kontani_PRL_2009,Morota_PRB_2011}. A similar dependence has also been found for the extrinsic SHE \cite{Tanaka_PTP_2012}.

The direct SHE and ISHE are not only intriguing from a fundamental physics point of view, but they also have become essential for many spintronics applications and devices. The ISHE is oftentimes used for a convenient spin-current detection in bilayers consisting of a ferromagnetic material close to a heavy metal. Prominent experimental examples are the combination of spin-pumping \cite{Saitoh_APL_2006,Kajiwara_Nat_2010,Mosendz_PRB_2010,Wang_PRL_2014,Zhang_PRL_2014,Jungfleisch_PRB_2015,Hahn_PRB_2013,Kelly_APL_2013,Wang_PRL_2014_anti,Du_JAP_2015,Sandweg_PRL_2011,Chumak_APL_2012,Jungfleisch_APL_2013,Jungfleisch_APL_2011,Zhang_PRB_2015,Weiler_PRL_2013,Weiler_PRL_2014} or
\newline
\begin{figurehere}
\centerline{
\includegraphics[width=0.82\columnwidth]{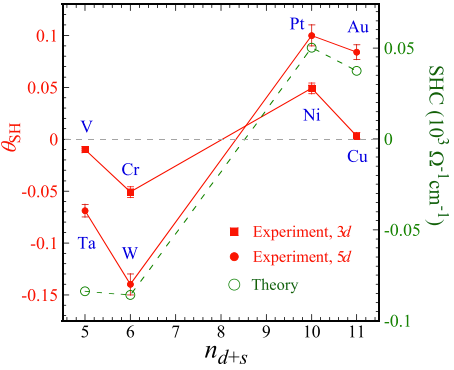}
}
\caption{
Spin Hall angles $\theta_\mathrm{SH}$ of $3d$ and $5d$ metals as a function of the total number of $d$ and $s$ electrons $n_\mathrm{d+s}$. The green open circles represent theoretical calculations of the spin Hall conductivity (SHC) for $5d$ metals and a good agreement, both in sign and magnitude, is found. Reprinted with permission from Ref.~[12]. Copyright 2015, AIP Publishing LLC.}
\label{fig1}
\end{figurehere} 
spin-Seebeck \cite{Uchida_Nat_2008,Uchida_Nat_2010,Uchida_APL_2010,Adachi_RPP_2013,Xiao_PRB_2010,Vlietstra_PRB_2014,Jungfleisch_APL_2013_heat,Kehlberger_PRL_2015,Agrawal_PRB_2014,Rezende_JMMM_2015,Rezende_PRB_2014,Schmid_PRL_2013,An_Nat_2013,Miao_PRL_2013,Qu_PRL_2013} experiments with the ISHE.

\subsection{Spin pumping}

Spin pumping describes the generation of a spin-polarized electron current in a normal metal in the vicinity of a ferromagnet that undergoes a resonant excitation (e.g., ferromagnetic resonance, excitation of spin waves) \cite{Tserkovnyak_PRL_2002,Czeschka_PRL_2011}. Spin pumping can be understood as a transfer of spin-angular momentum from a ferromagnet into a normal metal, thus reducing the magnetization. As a result of this process, the Gilbert damping of the magnetization precession is enhanced. This enhancement of the Gilbert damping parameter $\Delta\alpha$ is given by:
\begin{equation}
\Delta\alpha=\frac{g\mu_\mathrm{B}}{4\pi M_\mathrm{S}d_\mathrm{FM}}g^{\uparrow\downarrow}_\mathrm{eff},
\label{eq1}
\end{equation}
where $g$ is the $g$-factor, $\mu_\mathrm{B}$ is the Bohr magneton, $M_\mathrm{S}$ is the saturation magnetization, $d_\mathrm{FM}$ is the ferromagnetic film thickness, and $g^{\uparrow\downarrow}_\mathrm{eff}$ is the real part of the effective spin mixing conductance \cite{Tserkovnyak_PRL_2002}.

\subsection{Spin Seebeck effect}
The field of spin caloritronics \cite{Bauer_Nat_2012}, investigating the interplay between heat, charge and spin, has become a substantial research area in spintronics \cite{Uchida_Nat_2008,Uchida_Nat_2010,Uchida_APL_2010,Adachi_RPP_2013,Xiao_PRB_2010,Vlietstra_PRB_2014,Jungfleisch_APL_2013_heat,An_Nat_2013,Miao_PRL_2013,Qu_PRL_2013}. Analogous to the Seebeck effect where a temperature difference between two dissimilar electrical conductors or semiconductors gives rise to a voltage difference between the two materials, the spin Seebeck effect (SSE) describes the generation of a ``spin voltage'' when a magnetic material in contact with a normal metal is exposed to a temperature difference \cite{Uchida_Nat_2008,Uchida_Nat_2010}. Depending on the experimental configuration, the spin current generated by the SSE can either be perpendicular or parallel to the temperature gradient. The first arrangement is called transverse \cite{Uchida_Nat_2008}, the latter one longitudinal configuration \cite{Uchida_APL_2010}. The transverse SSE can be observed in both metals and insulators, whereas the longitudinal spin Seebeck effect can be detected unambiguously only in insulators because of the absence of the anomalous Nernst effect \cite{Schmid_PRL_2013}. The theoretical understanding of the SSE is still under a controversial debate. While some theories argue that the spin current results from the temperature difference between the ferromagnetic insulator and the metallic layer which results in a thermal interfacial spin pumping, e.g., Refs.~[\refcite{Uchida_APL_2010,Adachi_RPP_2013,Xiao_PRB_2010}], other authors propose that the longitudinal SSE originates from thermally-excited bulk magnons flowing across the thickness, e.g., Refs.~[\refcite{Kehlberger_PRL_2015,Agrawal_PRB_2014,Rezende_JMMM_2015,Rezende_PRB_2014}]. At the same time, it has recently been shown that SSE is not limited to ferromagnetically ordered systems, but can also be observed with antiferromagnets \cite{Wu_arxiv_2015} as well as paramagnets \cite{Wu_PRL_2015}.

Independent of the possible explanation for the spin-current generation in these heterostructures, the ISHE is generally used to detect the spin current.

\section{Route towards energy-efficient spintronics devices utilizing novel spin Hall detector materials}
\label{materials}

This Section is dedicated to investigations of novel, efficient spin Hall detector materials. First, we will review elementary materials (transition metals) as spin-to-charge current converters, followed by proximity induced ferromagnetically ordered Pt and Pd. Lastly, very recent studies utilizing antiferromagnets, materials showing magnetic ordering, but zero net magnetization, will be discussed.\newline
\begin{figurehere}
\centerline{
\includegraphics[width=0.95\columnwidth]{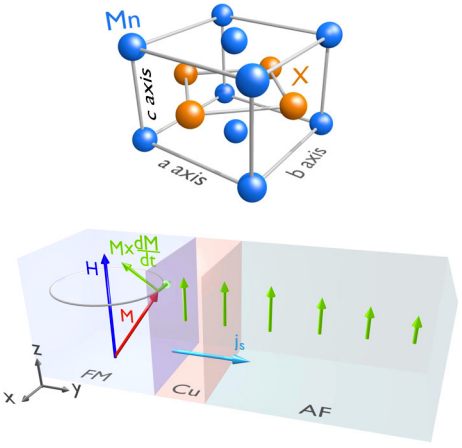}
}
\caption{Illustration of the chemical structure of CuAu-I-type antiferromagnets (X = Fe, Pd, Ir, Pt) and the spin pumping and spin Hall effect experiment for the Py/Cu/antiferromagnet structures. Reprinted figure with permission from Ref.~[7]. Copyright 2014 by the American Physical Society. 
}
\label{fig2}
\end{figurehere}
The inverse spin Hall effect in ferromagnet/Pt bilayers \cite{Saitoh_APL_2006,Kajiwara_Nat_2010,Mosendz_PRB_2010,Wang_PRL_2014,Zhang_PRL_2014,Jungfleisch_PRB_2015,Hahn_PRB_2013,Kelly_APL_2013,Uchida_Nat_2008,Hamadeh_PRL_2014} is widely used for spin current detection generated by either magnetization dynamics (e.g., ferromagnetic resonance) or a thermal gradient by the spin Seebeck effect \cite{Uchida_Nat_2008}. 
After the experimental discovery of the inverse spin Hall effect in ferromagnetic metal/normal metal (Ni$_\mathrm{80}$Fe$_\mathrm{20}$/Pt) bilayers \cite{Saitoh_APL_2006,Zhang_JAP_2015}, a variety of different metals and alloys were considered as potentially efficient spin Hall detectors. Mosendz et al. carried out combined spin pumping/ISHE measurements on Ni$_\mathrm{80}$Fe$_\mathrm{20}$/normal metal bilayers and reported spin Hall angles for Pt, Pd, Au, and Mo \cite{Mosendz_PRB_2010}. The analysis of these measurements for the spin Hall angle was later revised by Zhang et al. by determining the spin diffusion length of Pt through thickness dependent measurements \cite{Zhang_APL_2013}.

Furthermore, Wang et al. investigated bilayers of the ferrimagnetic insulator yttrium iron garnet (YIG) and various metals such as Cu, Ag, Ta, W, Pt, and Au with varying spin-orbit coupling strengths \cite{Wang_PRL_2014}. The spin Hall angles scale roughly as $Z^4$, where $Z$ is the atomic number, corroborating the spin-orbit coupling as underlying physical reason for the spin-charge interconversion. Figure~\ref{fig1} shows how the spin Hall angles $\theta_\mathrm{SH}$ of various $3d$ and $5d$ metals vary as a function of the total number of $d$ and $s$ electrons $n_\mathrm{d+s}$ \cite{Du_JAP_2015}. A good agreement with theoretical calculations of the spin Hall conductivity (SHC), both in sign and magnitude, is found \cite{Tanaka_PRB_2008}. This result also highlights the importance of the electron count of the $d$-bands \cite{Kontani_PRL_2009,Tanaka_PTP_2012,Tanaka_PRB_2008} and
confirms previous experimental investigations by Morota et al., where the spin Hall conductivities of the $4d$ and $5d$ transition metals Nb, Ta, Mo, Pd, and Pt were systematically studied by nonlocal spin valve measurements \cite{Morota_PRB_2011}.

Another pathway to design magnetic multilayers with optimal spin current efficiencies, is to study how magnetic ordering affects spin transport phenomena. In particular, it is interesting to address this question in widely used 
ferromagnetic metal/normal metal bilayers because many of the nominally non-magnetic materials are highly susceptible to magnetic proximity effects. Recently, it was shown that for Pt and Pd increased proximity induced magnetic moments can be correlated with strongly reduced spin-Hall conductivities \cite{Zhang_PRB_2015}. This observation is in agreement with the energy dependence of the intrinsic SHE determined by first principle calculations and can be understood, in a simple picture, as the development of a spin splitting of the chemical potential \cite{Zhang_PRB_2015}. 

Besides normally non-magnetic metals, which show proximity induced magnetic polarization, metals exhibiting magnetic ordering by themselves are interesting candidates as possible spin Hall detectors. In this context, antiferomagnets have recently attracted increased attention. In contrast to ferromagnets, antiferromagnets exhibit remarkable properties such as zero net magnetization, nontrivial spin-orbit coupling, and nonlinear magnetism. In addition, their excitations are at higher frequencies beyond ferromagnetic resonance. 
Due to their zero net magnetization additional ferromagnetic ordering is absent and, thus, other confounding effects of anisotropic magnetoresistance (AMR) can be ruled out. It was proposed theoretically that $\gamma$-FeMn, IrMn$_3$  and Cr show a large anomalous Hall effect and a SHE due to the large spin-orbit coupling and the Berry phase of the noncollinear spin textures in these antiferromagnets \cite{Shindou_PRL_2001,Chen_PRL_2014,Freimuth_PRL_2010}. \newline
\begin{figurehere}
\centerline{
\includegraphics[width=1\columnwidth]{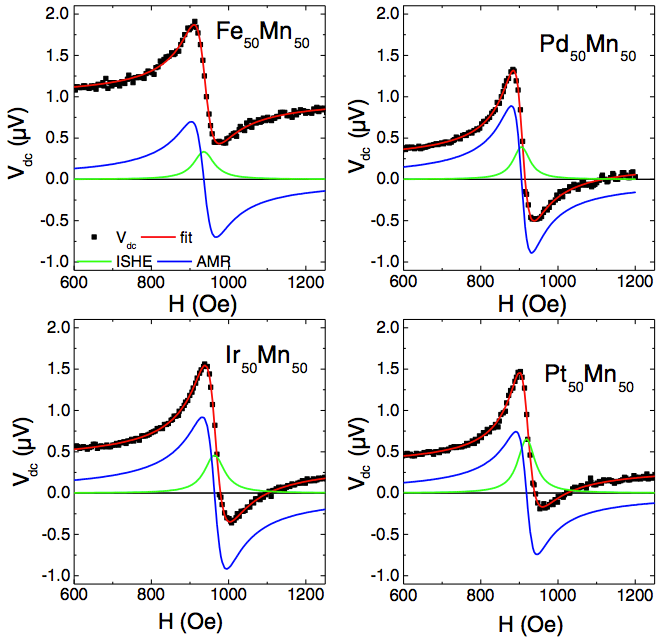}
}
\caption{Anisotropic magnetoresistance-inverse spin Hall effect spectra measured at 9 GHz of the Py(15~nm)/Cu(4~nm)/antiferromagnet(5~nm) structure for FeMn, PdMn, IrMn, and PtMn at room temperature. Reprinted figure with permission from Ref.~[7]. Copyright 2014 by the American Physical Society. 
}
\label{fig3}
\end{figurehere}

Mendes et al. investigated experimentally the antiferromagnetic metal Ir$_{20}$Mn$_{80}$, a material that is commonly used in spin-valve devices, as spin Hall material \cite{Mendes_PRB_2014}. In agreement with theoretical predictions, they found a large SHE as big as in Pt. Here, the spin-current injection in Ir$_{20}$Mn$_{80}$ is achieved by microwave spin pumping as well as by the longitudinal spin-Seebeck effect from a single crystal YIG. 

More systematic studies were reported by Zhang et al., where different CuAu-I-type metallic antiferromagnets were tested for their potential as spin-current detectors using spin pumping from a Ni$_{80}$Fe$_{20}$ layer and the inverse spin Hall effect \cite{Zhang_PRL_2014}. The investigated antiferromagnets feature the same chemical structure, i.e., X$_{50}$Mn$_{50}$ where X = Fe, Pd, Ir, and Pt (with increasing atomic number) as illustrated in Fig.~\ref{fig2}. By thickness-dependent measurements, the spin diffusion length of the investigated materials is found to to be all rather short, on the order of 1 nm. Figure~\ref{fig3} shows typical voltage spectra for the different types of investigated antiferromagnets. The estimated spin Hall angles of the four materials corroborate the importance of spin-orbit coupling of the heavy metals for the properties of the Mn-based alloys through orbital hybridization \cite{Zhang_PRL_2014}. First-principles calculations of ordered alloys showed that the value of the spin Hall conductivity depends strongly on the crystal orientation and staggered antiferromagnetic magnetization. A follow-up work by Zhang and co-workers indeed showed a growth-orientation dependence of spin torques by studying epitaxial samples \cite{Zhang_APL_2011,Zhang_PRB_2013}, which may be correlated to the anisotropy of the SHE \cite{Zhang_arxiv_2015}.

{Besides being promising materials for spin current generation and detection, some antiferromagnetic materials also appear to be good candidates for transmitting spin currents. It was shown that spin currents can be injected efficiently from a magnetic insulator (YIG) into an antiferromagnetic insulator (NiO) \cite{Wang_PRL_2014_anti}. The insertion of a thin NiO layer between YIG and Pt enhances the spin-current injection into Pt, which suggests a high spin-transfer efficiency at both YIG/NiO and NiO/Pt interfaces and spin transport in NiO mediated by antiferromagnetic magnons or antiferromagnetic fluctuations \cite{Wang_PRL_2014_anti}.}

\begin{figurehere}
\centerline{
\includegraphics[width=1\columnwidth]{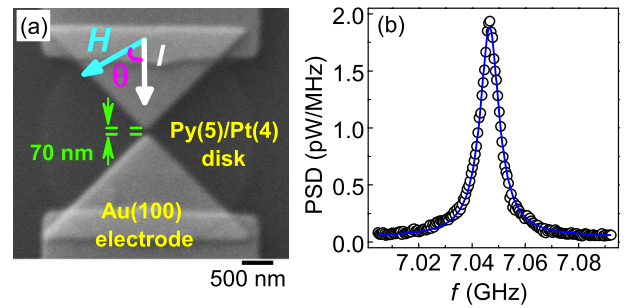}
}
\caption{(a) Scanning electron microscope image of spin-torque nano-oscillator. (b) Power spectral density of the emitted microwave signal at $H=700$~Oe, $I=20$~mA, $T=6$~K and at an angle $\theta=60^\circ$. Reprinted figure with permission from Ref.~[72]. Copyright 2013 by the American Physical Society. 
}
\label{fig4}
\end{figurehere}

\section{New approaches to drive collective magnetization dynamics by spin Hall effect}
\label{collective}
Magnetization dynamics is conventionally driven at rf frequencies by the Oersted field generated by a microwave signal. However, over the past decades new methods were developed using dc current induced spin currents to exert spin torques on the magnetization. In this Section, we will give a very brief overview over the more traditional ways to excite spin dynamics by dc currents and review alternative approaches on basis of SHEs reported very recently.

Pioneering work by Slonczewski and Berger predicted that a flow of spin angular momentum can exert a spin-transfer torque \cite{Haney_PRB_2013} on the magnetization of a ferromagnet and drive it out of equilibrium \cite{Slonczewski_JMMM_1996,Berger_PRB_1996}. This spin torque can act as a negative magnetic damping and even excite auto-oscillations of the magnetization \cite{Kiselev_Nature_2003,Slavin_IEEE_2009}. Spin-torque nano-oscillators have been demonstrated in nanopillars and spin valves \cite{Silva_JMMM_2010,Mohseni_Science_2013}, point contacts \cite{Ozyilmaz_PRL_2004} and magnetic tunnel junctions \cite{Deac_Nat_2008}. The basic idea of all these structures is to generate a spin polarization of the current that is noncollinear with the background magnetization. Usually, this is achieved by patterning of multilayered films. 
These multilayered films consist of a ``reference layer'' acting as a polarizer that spin-polarizes an electric charge current and a ``free layer''. If the reference and free layer magnetizations are noncollinear, the spin-polarized current will destabilize the magnetization in the free layer by $sd$-exchange interaction and eventually initiate auto-oscillations in the free layer when a certain threshold current is reached.

Recently, a new type of spin-torque nano-oscillators based on spin-orbit torques from a charge current was realized in Ni$_{80}$Fe$_{20}$/Pt bilayers \cite{Demidov_Nat_2012,Liu_PRL_2013,Demidov_SciRep_2015}. 
Here, the spin current is generated by means of the SHE which makes simplified device structures without complicated multilayer stacks possible. Figure~\ref{fig4}(a) shows a possible realization of spin Hall nano-oscillator, which comprises a $4~\mu$m Py(5~nm)/Pt(4~nm) disk and Au electrodes with a separation of 70~nm. A typical power spectral density spectrum of the emitted microwave signal is shown in Fig.~\ref{fig4}(b). Even more important than the simpler device geometry is that the utilization of the SHE allows the realization of magnetic nano-oscillators based on conducting {\it and} insulating materials that do not require charge current flow through the active device area \cite{Demidov_Nat_2012}.

First approaches to drive auto-oscillations by a dc current in bilayers of the ferrimagnetic insulator YIG and Pt did not succeed because the estimated threshold current that is required to trigger the self-oscillations exceeded the experimental feasibility \cite{Hahn_PRB_2013,Kelly_APL_2013}. These experiments were performed on large, macroscopic structures, where many nearly degenerate spin-wave modes are present. When the dc source is applied, all of these modes are driven simultaneously and, thus, they compete with each other. This leads to a self-limitation since none of these modes can overcome the threshold for the onset of auto-oscillations \cite{Hamadeh_PRL_2014,Collet_arxiv_2015}. In order to overcome these difficulties, it is crucial to reduce the density of the mode spectrum, and, ideally, to isolate one single mode \cite{Jungfleisch_PRB_2015,Hamadeh_PRL_2014}. This discretization can be achieved by reducing either the thickness \cite{Jungfleisch_PRB_2015} and/or the lateral dimensions \cite{Hamadeh_PRL_2014,Collet_arxiv_2015} of the YIG pattern, because quantization leads to increased frequency gaps between the modes. For this purpose, nanometer-thin, ultra-low damping YIG films are indispensable \cite{Jungfleisch_JAP_2015,Chang_IEEE_2014,Yu_SciRep_2014}. 

Hamadeh et al. use magnetic resonance force microscopy (MRFM) to show damping compensation in 5~$\mu$m diameter YIG(20 nm)/Pt(7 nm) disks \cite{Hamadeh_PRL_2014}. They demonstrate that the magnetic losses of spin-wave modes existing in the magnetic insulator can be reduced or enhanced by a factor of 5 depending on the polarity and magnitude of an in-plane dc current flowing through the adjacent normal metal with strong spin-orbit interaction. Figure~\ref{fig5}(a) illustrates a density plot of MRFM spectra as a function of field and current. The signal is asymmetric in the applied dc current. The signal broadens and the amplitude decreases as the current is increased to positive values and almost disappears at +8~mA. For negative currents, it becomes narrower and the amplitude is maximal for currents smaller than -10~mA. The normalized integrated power increases by a factor of 5 from +12~mA to -12~mA, see Fig.~\ref{fig5}(b). As is apparent from Fig.~\ref{fig5}(c), Hamadeh and co-workers observe an increase of the linewidth from 6~Oe at 0~mA up to 14~Oe at 12~mA and they find a minimum of about 2~Oe between -8 and -11~mA.  In order to check if this current can be identified as the threshold current for the onset of self-oscillations, they show MRFM spectra in an experiment where no rf excitation is applied [Fig.~\ref{fig5}(d)]. Strikingly, the narrow peak shifts linearly in dc current with the applied field,
\begin{figurehere}
\centerline{
\includegraphics[width=1\columnwidth]{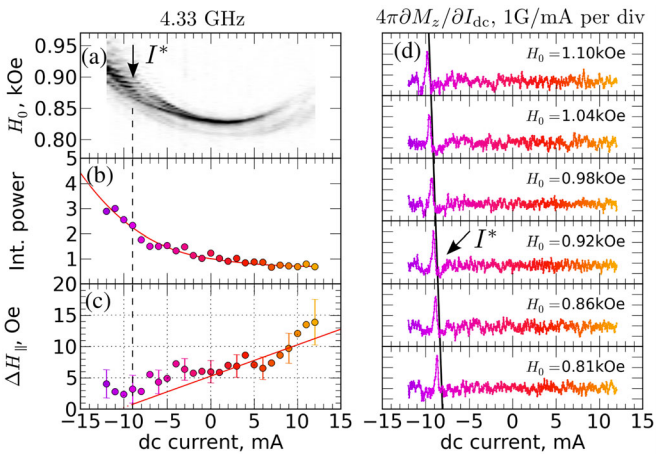}
}
\caption{(a) Density plot of the MRFM spectra at 4.33 GHz as function of field and current. The
color scale represents the change in the magnetization (white: 0 G, black: 1.5 G). (b) Integrated power versus applied current. (c) Dependence of linewidth on current. (d) Differential measurements modulated by 0.15 mA, no rf excitation, versus current at six different values of the in-plane magnetic field. Reprinted figure with permission from Ref.~[51]. Copyright 2014 by the American Physical Society. 
}
\label{fig5}
\end{figurehere}
which is in agreement with the theoretically expected behavior for auto-oscillations.

Another interesting approach to drive uniform magnetization dynamics is spin-torque ferromagnetic resonance (ST-FMR) which was originally developed for all-metallic systems \cite{Liu_PRL_2011}. The basic idea is as follows: If a flow of alternating charge current is passed through a heavy normal metal/metallic ferromagnet bilayer, the SHE generates an oscillating transverse spin current in the normal metal. This results in a spin accumulation at the interface and, thus, it leads to a transfer of spin angular momentum to the ferromagnetic layer that can induce ferromagnetic resonance. In metallic ferromagnets, the resonance is detected by a homodyne voltage from anisotropic magnetoresistance \cite{Gui_PRL_2007,Bai_PRL_2013}. The ST-FMR technique allows for a convenient determination of the spin Hall angle by a simple lineshape analysis, which is in agreement with measurements of the dc-current dependence of the resonance linewidth \cite{Liu_PRL_2011}. Besides investigations on Py/Pt bilayers, ST-FMR metrology was also used in ferromagnet(Py)/topological insulator(Bi$_2$Se$_3$) systems \cite{Mellnik_Nature_2014} and at a ferromagnet(Py)/Rashba-interface(Ag/Bi) \cite{Jungfleisch_arxiv_2015_Rashba}.

Recently, it was predicted that the concept of ST-FMR, where a ac spin Hall effect mediated spin accumulation  drives magnetization oscillations, can be extended to insulating systems \cite{Chiba_PRA_2014,Chiba_JAP_2015}. Here, we have to consider two different contributions to the measured voltage signal: (1) spin pumping and (2) spin Hall magnetoresistance (SMR) \cite{Nakayama_PRL_2013}. In these kind of magnetic insulator/normal metal bilayers, SMR mixes with the microwave signal rather than the anisotropic magnetoresistance. SMR stems from the spin-current backflow at a ferromagnetic insulator/normal metal interface: If a charge current is passed through the normal metal, it is converted into a spin current that accumulates at the interface to the ferromagnetic insulator. If the spin polarization $\sigma$ and the magnetization $M$ are noncollinear, spin-flip scattering can occur. This leads to a partial absorption of the spin current at the interface which results in the excitation of magnetization precession.\newline

\begin{figurehere}
\centerline{
\includegraphics[width=0.9\columnwidth]{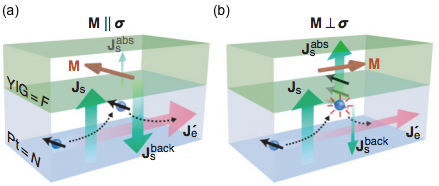}
}
\caption{(a) and (b) Geometric relation between the flow of electrons and accumulated spins in the normal metal and the magnetization in the magnetic insulator. 
Reprinted figure with permission from Ref.~[85]. Copyright 2013 by the American Physical Society. 
}
\label{fig6}
\end{figurehere}

Consequently, less spin current is back-reflected. This absorption is maximized when $M$ is perpendicular to $\sigma$ and zero when $M$ is parallel to $\sigma$ \cite{Nakayama_PRL_2013}, see Fig.~\ref{fig6}. Thus,  the conductivity enhancement due to SHE and ISHE is maximized (minimized) when $M$ is perpendicular (parallel) to the charge 
current $J_\mathrm{e}$, because $J_\mathrm{e}$ is perpendicular to $\sigma$ and the Pt resistance depends on the magnetization direction in the YIG. {The basic idea of SMR was already proposed by D'yakonov in 2007 \cite{Dyakonov_PRL_2007}. Here, an externally applied magnetic field destroys the edge spin polarization resulting in a positive magnetoresistance. In the work by Nakayama et al. exchange interaction dephasing the spin accumulation is the reason for the observed magnetoresistance.} 

Chiba et al. developed a model of the ac spin Hall magnetoresistance in a bilayer system consisting of a magnetic insulator and a heavy metal \cite{Chiba_PRA_2014,Chiba_JAP_2015}. As is shown in Refs.~[\refcite{Chiba_PRA_2014,Chiba_JAP_2015}], it is possible to derive expressions for dc voltages under ST-FMR in these systems using the drift-diffusion spin model and quantum mechanical boundary condition at the interface. 
Figure~\ref{fig7} shows the calculated dc voltage spectra according to their model. The inset illustrates that the spin-pumping contribution is purely symmetric and the SMR features a superimposed symmetric and antisymmetric Lorentzian lineshape. Very recent experimental studies show that the concept of ST-FMR can indeed be applied to insulating systems \cite{Schreier_arxiv_2014,Sklenar_arxiv_2015,Jungfleisch_arxiv_2015}.
\newline

\begin{figurehere}
\centerline{
\includegraphics[width=0.9\columnwidth]{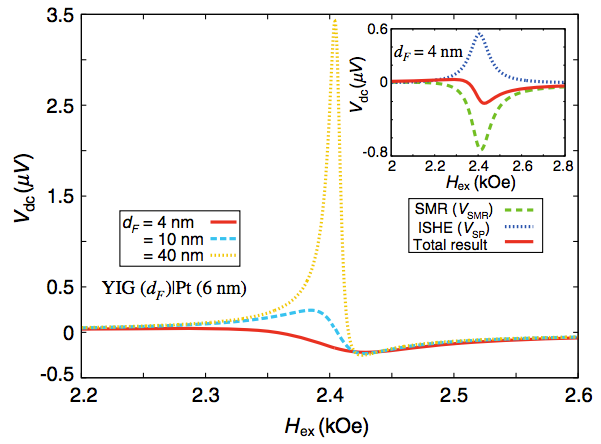}
}
\caption{Calculated dependence of the ST-FMR spectra on the ferromagnetic layer thickness $d_\mathrm{F}$ at a driving frequency of 9 GHz and an in-plane angle of 45$^\circ$. Inset: Contributions by a rectification due to spin Hall magnetoresistance and spin pumping. Reprinted figure with permission from Ref.~[83]. Copyright 2014 by the American Physical Society. 
}
\label{fig7}
\end{figurehere}

\section{Manipulation of spin textures by spin Hall effects}
\label{textures}
In this Section, we will provide a short summary of different, widely known spin textures and introduce novel ideas, which make use of SHEs to dynamically create and manipulate magnetic skyrmion bubbles.

Competing magnetic interactions, most importantly exchange interactions at short length scales and magnetostatic dipole interactions at long length scales, result in the formation of a variety of magnetization states and spin textures resulting in many fascinating physical phenomena. These spin textures exhibit technologically relevant features including emergent electromagnetic fields and efficient manipulation \cite{Jonietz_Science_2010,Yu_Nat_Mater_2011,Nagaosa_Nat_Nano_2013,Lin_PRL_2013,Zang_PRL_2011,Jiang_Science}.

One of the easiest examples are magnetic domain walls, the boundaries between magnetic domains with different magnetization configurations. Magnetic domain walls exhibit interesting physical phenomena including complex dynamics at different time scales  \cite{Novosad_PRB_2005,Choe_Science_2004}; typical domain wall widths lie in the nanometer-range. Another prominent example of spin textures are magnetic vortices that are stabilized by magnetostatic interactions. Magnetic vortices are planar spin textures with an in-plane magnetization curling leading to two possible chiralities and a central singularity that is called vortex core. A vortex core points perpendicular to the film plane and can thus have two polarities. They have typically a diameter of a few nm \cite{Wachowiak_Science_2002}. \newline


\begin{figurehere}
\centerline{
\includegraphics[width=0.9\columnwidth]{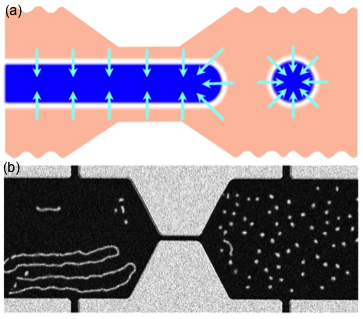}}
\caption{(a) Illustration of the formation of magnetic skyrmions from chiral stripe domains through a geometrical constriction. (b) Kerr Microscopy image taken after applying a short current pulse, showing a large number of skyrmions formed on the right side of the constriction. From Ref.~[95]. Reprinted with permission from AAAS.} 
\label{fig8}
\end{figurehere}

Very recently, a similar, but more complex type of topologically stable spin textures has gained great attention in the scientific community: magnetic skyrmions. They were first observed by neutron scattering at the border between paramagnetism and long-range helimagnetic order perpendicular to small applied magnetic fields \cite{Skyrmion_Science_09}. The novelty of magnetic skyrmions and what sets them apart from the above mentioned spin textures is that they are three-dimensional in character. There are two types of magnetic skyrmions: (1) ``hedgehog skyrmions'', where the progression of the magnetization is cycloidal across the diameter and (2) ``vortex skyrmions'', where the progression of the magnetization is helical. The ability to controllably create and move magnetic skyrmions is of fundamental importance for technological implementation of skyrmion-based spintronics. Here, the stability of skyrmions at room temperature is key and has been challenging in the past. Recently, it was shown that 
magnetostatically stabilized skyrmion structures, magnetic bubbles, can form in magnetic thin films with perpendicular magnetic anisotropy at room temperature \cite{Jiang_Science}. The electric-current generation of skyrmions is achieved by adding an additional layer with strong spin-orbit coupling to the ferromagnet which results in an interfacial broken inversion symmetry and, thus, the generation of interfacial Dzyaloshinskii Moriya interaction (DMI). DMI stabilizes chiral magnetic domain walls around the bubble that lead to a skyrmion spin structure. These skyrmion bubbles can then be electrically manipulated by SHEs. Jiang et al. demonstrated this in a Ta/CoFeB/TaO$_\mathrm{x}$ trilayer, where skyrmions can be generated via laterally inhomogeneous current-induced spin-orbit torques in a process analog to the droplet formation in surface-tension driven fluid flow \cite{Jiang_Science}.

In the Ta/CoFeB/TaO$_\mathrm{x}$ heterostructure the electric current flowing through the heavy metal generates a transverse spin current due to the SHE, which results in spin accumulation at the interface with the ferromagnetic layer \cite{Jiang_Science}. 
This spin accumulation exerts a spin-orbital torque on the chiral domain wall. If the current flow is homogeneous, a chiral spin-orbital torque enables efficient domain-wall motion \cite{Emori_Nat_2013,Ryu_Nat_2013}. 
Because of symmetry reasons the torques on a stripe domain with chiral walls cancel on the sides parallel to the current and, thus, only the end of the stripe domain is moved. For a pinned domain, this results in an elongation of the stripe. However, by introducing a geometrical constriction into the current-carrying trilayer wire, a transverse current component around the narrow neck can be achieved. 

As a result an inhomogeneous effective force caused by the spin Hall field is created and extends the end on the domain. The surface tension in the domain wall increases because of the continuously expanding radius, which finally results in breaking the stripes into circular domains [see Fig.~\ref{fig8}(a)]. Due to the presence of interfacial DMI, the circular domains maintain a well-defined chirality and once formed, these created synthetic hedgehog (N\'eel) skyrmions are stable due to topological protection. They can be moved very efficiently in the direction of the charge current, Fig.~\ref{fig8}(b). Depending on the strength of the external magnetic field, these dynamically created skyrmions feature a variable size between $700$~nm and 2~mm and are stable for at least 8 hours \cite{Jiang_Science}. Figure~\ref{fig9} shows an experimentally determined electric current vs. magnetic field phase diagram for the skyrmion formation. In principle, the size of the skyrmions could be scaled down \cite{Woo_arxiv_2015,Moreau-Luchaire_arxiv_2015} by engineering  material-specific parameters that control the various competing magnetic interactions which might lead to sophisticated  skyrmionic device concepts such as a spin Hall controlled skyrmion racetrack memory \cite{Jiang_Science}. \newline

\begin{figurehere}
\centerline{
\includegraphics[width=0.9\columnwidth]{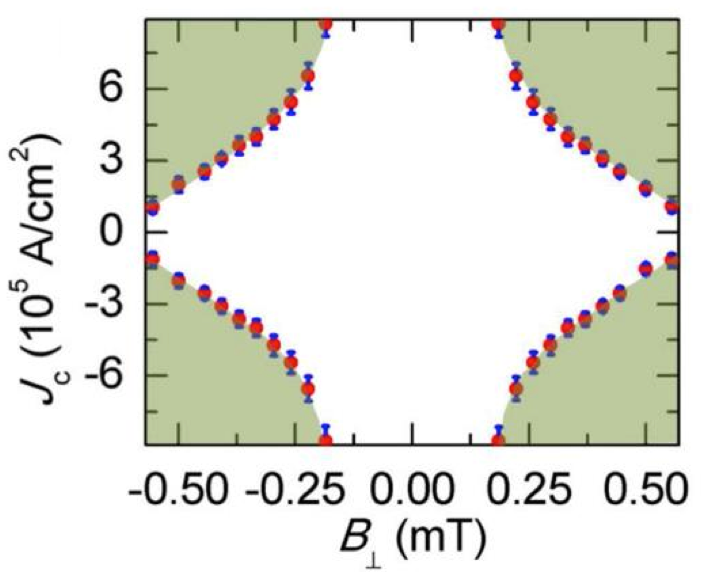}}
\caption{Phase diagram for skyrmion formation. The shaded area illustrates field-current combinations that result in the persistent generation of skyrmions after each current pulse. From Ref.~[95]. Reprinted with permission from AAAS.}
\label{fig9}
\end{figurehere}

As compared to the extensively studied magnetic domains and domain walls in the metallic systems, another interesting, but yet missing aspect is the electrical control of magnetic domains and domain walls in magnetic insulators such as YIG \cite{Jiang_PRL_2013}. This is made possible by the SHE in close contact to the magnetic insulator. Historically, it is well known that rare-earth doped YIG films contain periodic magnetic bubble domains \cite{Slonczewski_book_1979}. Figure~\ref{fig10} shows such periodic bubble domains in a 6~$\mu$m thick Bi doped YIG film. It may be that these magnetic bubbles in YIG films are topologically different from coherent skyrmion spin textures in $B20$ compounds, due to the absence of DMI. By employing the SHE of various heavy metals in heavy metal/YIG hybrids, it would be intriguing to see the motion of these bubbles. These investigations could shed light on the question if the direction of bubble motion may serve as a criterion to select the bubbles with different skyrmion numbers. On the other hand, as compared to lithographically patterned micro-disks of a magnetic insulator, it is also possible that these naturally existing bubbles can be driven into auto-oscillation by means of SHE and spin-transfer torque. \newline

\begin{figurehere}
\centerline{
\includegraphics[width=0.9\columnwidth]{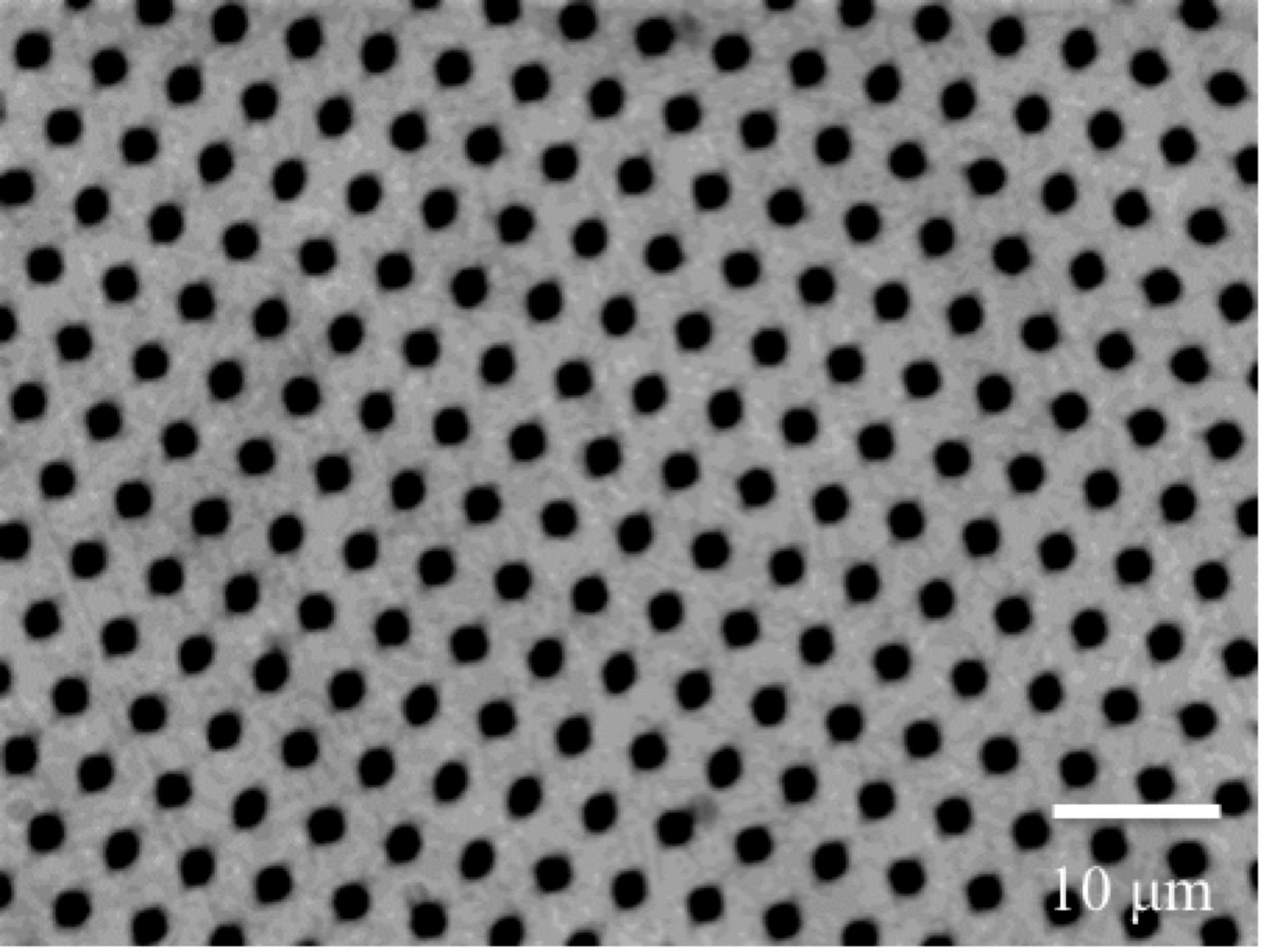}}
\caption{Polar magneto-optical Kerr effect image of periodic magnetic bubble domains in the 6~$\mu$m thick Bi doped YIG film grown on GGG (111) substrate by liquid phase epitaxy. The diameter of each magnetic bubbles is around 1.3~$\mu$m.}
\label{fig10}
\end{figurehere}

\section{Conclusion}

More than two decades after the discovery of the giant magnetoresistance, spintronics is still a vivid field of contemporary magnetism research that constantly grows and develops new ideas. In particular, it gained new momentum in recent years after the experimental demonstration of many spin Hall effect related effects such as the spin Seebeck effect, detection of spin-pumping driven spin currents or spin-torque ferromagnetic resonance to name only a few.

Predicting future research directions is certainly an impossible task, but we want to give a brief outlook on possible pathways to novel spin Hall effect related spintronics applications. 

The search for optimal spin Hall detector materials has definitely not yet come to an end. The utilization of antiferromagnets in spintronics devices is a promising route; in particular taking into account their high-frequency characteristics beyond ferromagnetic resonance, which makes them interesting for information technologies. Besides that topological insulators \cite{Mellnik_Nature_2014,Fan_Nat_2014} might revolutionize the efficiency charge-spin current interconversion, and lead to new spin Hall driven spin-torque devices.

The implementation of magnetic insulators in spintronics has the potential for the development of novel low-energy consuming devices. Many of the above mentioned effects were first shown in all-metallic systems, and later on in insulators. Along those lines, it would be very interesting to demonstrate the movement of skyrmions bubbles in insulators by spin Hall effect driven spin torques or even the onset of auto-oscillations.

Beyond bulk effects such as spin Hall effects, one might think about the utilization of interface- or surface effects such as the Rashba-Edelstein effect (REE). First experimental studies show that REEs can indeed be used for the spin- to charge-current interconversion \cite{Jungfleisch_arxiv_2015_Rashba,Sanchez_Nat_2013,Zhang_JAP_2015_Rashba}.

Many surprises can be expected in this rapidly evolving field and we look forward to exciting discoveries in the future.

\nonumsection{Acknowledgments} \noindent This work was supported by the U.S. Department of Energy,
Office of Science, Materials Science and Engineering Division. 

\nonumsection{Preprint of an article published in SPIN 05, 1530005 (2015) [13 pages] DOI: 10.1142/S2010324715300054 \copyright \, copyright World Scientific Publishing Company, \url{http://www.worldscientific.com/worldscinet/spin}.}

\end{multicols}
\end{document}